\documentclass[pdflatex,sn-mathphys-num, iicol]{sn-jnl}

\usepackage{graphicx}%
\usepackage{multirow}%
\usepackage{amsmath,amssymb,amsfonts}%
\usepackage{amsthm}%
\usepackage{mathrsfs}%
\usepackage[title]{appendix}%
\usepackage{xcolor}%
\usepackage{textcomp}%
\usepackage{manyfoot}%
\usepackage{booktabs}%
\usepackage{algorithm}%
\usepackage{algorithmicx}%
\usepackage{algpseudocode}%
\usepackage{listings}%

\theoremstyle{thmstyleone}%
\theoremstyle{thmstyletwo}%

\theoremstyle{thmstylethree}%

\begin{document}

\title[Track fitting as a language translation problem]{Track fitting as a language translation problem}


\author*[1]{\fnm{Deepak} \sur{Samuel}}\email{deepaksamuel@cuk.ac.in} 

\author[2]{\fnm{Christy} \sur{Elsa Koshy}}\email{christyelsa62@gmail.com}


\affil*[1,2]{\orgdiv{Department of Physics}, \orgname{Central University of Karnataka}, \orgaddress{\street{Aland Road}, \city{Kalaburagi}, \postcode{585367}, \state{Karnataka}, \country{India}}}




\abstract{

Particle physics analysis depends on information retrieved from particle tracks in a detector. Traditionally, this involves collecting individual detector hits, recombining them to form tracks, and fitting them to a model function. Large Language Models (LLMs) have demonstrated remarkable proficiency in comprehending the grammar and semantics of languages. Our exploratory study leverages transformers to conceptualise particle tracking as a language translation problem, wherein the language of detectors is translated into the language of physics. A simple decoder-only architecture with a custom tokenizer was designed for this task, and the behaviour of the model in this atypical setting provided interesting insights. We evaluated the advantages of this formulation in terms of track-parameter accuracy, and the ease of representing multi-track events, and finally tested the model on real cosmic-muon data. On synthetic datasets, the model outperformed a simple regression technique by an order of magnitude in track-parameter accuracy while being on par with the random sample consensus (RANSAC) method. The model also identified multi-track events with an accuracy of 98\%. The analysis of real cosmic-muon events showed an excellent agreement of the zenith angle distribution with expectations. Our results suggest that a compact language model shows competitive performance and a natural ability to represent multi-track events. Further studies with complex geometries are required to understand the full potential of this technique.
}
\maketitle
\section{Introduction}Particle physics experiments generally involve collisions or interactions of particles and analysing the tracks of outgoing particles through a data analysis pipeline. 
For instance, the proposed Iron Calorimeter (ICAL) detector at the India-based Neutrino Observatory is designed to detect muons produced from the charged-current interactions of atmospheric neutrinos with Iron nuclei \cite{mondal2012india}. The muon tracks have encoded in them, the energy, momentum and incident angle of the parent neutrinos, offering a wealth of information about the origin and nature of neutrinos. The muon track and its associated kinematical parameters are reconstructed from the spatial coordinates of detector hits recorded in active detector layers through a process called track reconstruction.

Track reconstruction begins with the collection and isolation of hits produced by individual muons from noise hits or possibly other additional particles (including other muons) that are produced in the interaction. Typically, for particles in a magnetic field, a Hough transform is performed to identify tracks that follow a given topology.  Such tracks identified through the track-finding process are then passed to track-fitting algorithms, such as Kalman filters, to extract the kinematical parameters \cite{Bhattacharya2014}. In certain experiments, particularly those involving high pile-up, the computational overhead of track-finding becomes a bottleneck within the analysis chain, necessitating the development of fast algorithms on GPU and machine learning models \cite{Chetri2026}\cite{choma2020track}. 

Driven by the success of LLMs in various domains, specifically code generation and translation, we seek in this study to formulate track reconstruction as a language translation problem. In translation tasks, a sequence of tokens from one language is mapped to another language. Similarly, the objective of track-fitting is to map a sequence of detector hits to track parameters. Both these tasks, therefore, share a common skeleton of parsing a set of input tokens and contextualising them to generate another set of tokens, without changing the meaning of the original statement. Specifically, in translating from one language to another, while the grammar and semantics change, the information content should remain invariant. 

Transformer architectures are a natural fit for sequence-to-sequence tasks \cite{Vaswani2017}. However, a natural question arises regarding the semantics of this new language. Considering language as a medium of human thought makes its description a creative endeavour. Therefore, while many descriptions may serve this purpose, motivated by the work of Yoran et.al, we decided to express the track parameters as part of a computer programme, more specifically, a pre-defined Python function \cite{yoran_kolmogorov_2025}. In their study, an LLM was trained to generate the shortest possible code to generate random sequences from DNA sequences, audio, and text. It was shown that even flagship models performed poorly, whereas a custom model trained on a domain-specific language that consists of a limited set of instructions faired well with synthetic data but still failed with real datasets. In our study, a particular advantage arises from the sequences being determined by the well-defined trajectory of the particles, rather than being entirely random.

Similar to a language that follows grammatical constraints, code generation also follows syntactic and numerical constraints, closely aligning with our objective. 
By choosing an appropriate syntax for the function, one can also perform signal-noise separation, allowing us to leverage the robustness of LLMs to malformed inputs.

In this novel study, we aimed to verify whether the effectiveness of LLMs in translation and code generation tasks transfers to a domain that requires more complex numerical reasoning and combinatorial signal-background discrimination. Previous studies have shown that even state-of-the-art LLMs exhibit weaknesses in their numerical reasoning capabilities \cite{li-etal-2025-exposing}.  Therefore, the motivation for this study was purely exploratory, and we do not claim, at this stage, the superiority of our technique over any existing methods.

The cosmic muon detector at TIFR, Mumbai, served as a test bench for this study, significantly simplifying the analysis because of the straightforward linear trajectories of the particles. This detector was a prototype version of the ICAL detector of the India-based Neutrino Observatory. We begin with a brief description of the detector geometry and then introduce the grammar of the new language. The details of the model and the training conditions are presented, followed by the final results.

\section{Detector geometry}
\begin{figure}
    \centering
    \includegraphics[page=1, trim=5cm 6cm 5cm 2cm, clip, width=1\linewidth]{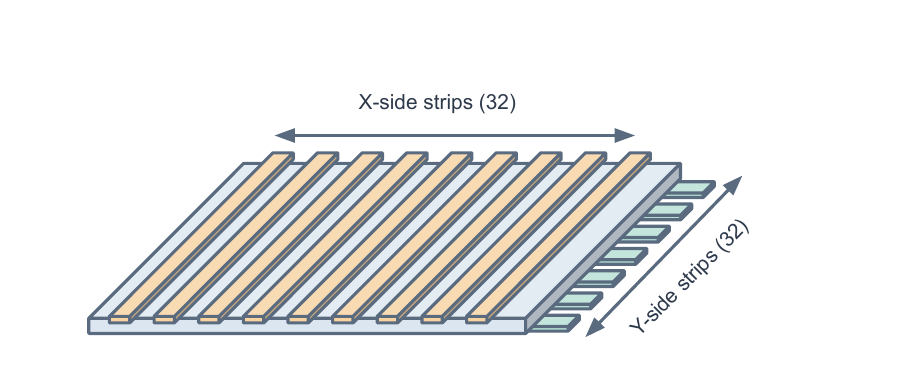}
    \caption{Schematic diagram of a Resistive Plate Chamber showing the strip placements on the top ($x$-side) and at the bottom ($y$-side). In the prototype detector, there are 32 strips on either planes. The strip numbers on theses planes encode the $x$, $y$ coordinates of the particle passage through these detectors. 12 such RPCs are stacked on top of each other with a gap in between each one. The location of the RPCs along the height of the stack gives the $z$-coordinate.}
    \label{fig:rpc}
\end{figure}

\begin{figure}
    \centering
    \includegraphics[width=1\linewidth]{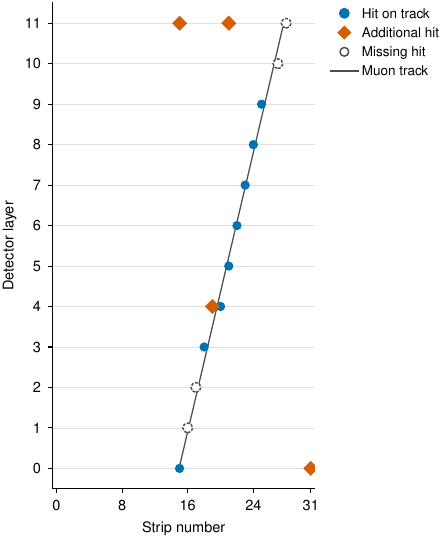}
    \caption{A sample cosmic muon event showing the hits on the individual RPCs including additional hits due to detector noise. The hollow circles represent the muon coordinate in layers that did not respond to a passing muon due to detector inefficiencies.}
    \label{fig:muon_track}
\end{figure}

The prototype detector at TIFR, Mumbai consists of  12-layer Resistive Plate Chambers (RPCs) each with 32 strips on the top and bottom planes as described in Figure \ref{fig:rpc} \cite{DATAR2009744}. The RPCs are of dimensions 1 m $\times$ 1 m with the strip width roughly 3 cm. The RPCs are made of two glass plates separated by a small gap in which a suitable gas mixture that is ionised by passing muons is circulated.  With a gap of 16.8 cm between adjacent RPCs, the total height of the stack is about 1.85 m. When a muon passes through a RPCs, the specific  $x$ and $y$ strips through which it passed give rise to an electrical signal. Thus the strip numbers in a particular layer encode the $x$ and $y$ coordinates of the muon in that layer. When a muon crosses multiple layers, a coincidence trigger can be used as a signal for the data acquisition to record the strip hits from all the layers. Using these strip number from each layers, one can estimate the track of the muon which then can be used to perform other physics analysis. A sample cosmic muon event is shown in Figure \ref{fig:muon_track}.

The aforementioned description is sufficient for the discussions that follow, and interested readers are referred to other studies that describe the detector and data acquisition in detail \cite{bhuyan_vme-based_2012}.

\section{Data format and track fitting}
On receipt of a trigger signal, the data acquisition stores the strip hits and the timing from all the layers in a ROOT file. For the purposes of this study, we omit the timing signal and consider only the spatial coordinates. Also, since most of analysis was performed using python codes, the data in the ROOT files were converted in plain text format. The strip hit information of strip $s$ that was hit in  layer $l$ is encoded in these text files as:

\begin{equation*}
  \textbf{Hit ID}=  l \times 32 + s
\end{equation*}

A track is therefore represented by an array of numbers:
\begin{equation*}
    [15, 49, 83, 117, 151, 185, 219, 253, 287, 321, 355]
    \label{eqn:array}
\end{equation*}
If one ignores the multiple couloumb scattering effects in the RPCs and in the support structures, the trajectory of cosmic muons is roughly a straight line. Therefore, track fitting in this context refers to finding the slope and intercept of the cosmic muon track.

As shown in Figure \ref{fig:muon_track}, additional hits due to noise may be present in addition to the muon hits in an event. These additional hits may also be due to the size of the ionisation cluster in the gas, which  leads to hits in the adjacent strips, as in layer 4 in Figure \ref{fig:muon_track}.  Furthermore, detector inefficiencies may cause a layer not to respond to a cosmic muon. Previous studies have shown that the average strip hit multiplicity is approximately 1.5 strips, and the efficiency is approximately 90\%.

Therefore, the quality of the fit is impaired by the presence of outliers and detector inefficiency.

\section{Grammar and semantics of our language}
The goal of track reconstruction is to retrieve from detector hits, the track parameters, which, in turn, will be used for downstream physics analysis. In the prototype detector, the track parameters are the slope and intercept. These parameters can be used, for instance, for analysis of the arrival direction of the cosmic muons, specifically, the zenith angle $\theta$ and azimuth $\phi$ \cite{pal_measurement_2012}.
To rephrase a track fitting problem to a language translation problem, the description of the new language is to be first decided. Whatever be the description of the new language, one requirement that needs to be satisfied is that no information should be lost in translation. In other words, one should be able to re-translate it into the original language and return to the original description.
The original description is in the form of an array of integers which encode the positional information of the muons as they propagate through the detector.
The new language intended to replace track fitting should not only include track parameters but also accommodate noise and other detector effects.

One approach is to represent the track parameters as arguments of a function such that when the same is executed, the hit-IDs are returned as shown below:
\begin{equation*}
    [15,49,83,117,151,...] \leftrightarrow  gen\_evt(slp=[2], icpt=[15])
    \label{eqn:arraytocode}
\end{equation*}
The $slp$ and $icpt$ represent the track parameters, specifically the slope and intercept. They are represented as arrays to allow representation of multiple tracks.

Building on this idea, noise hits and other effects can be represented as additional arguments to this function.
\begin{equation*}
   gen\_evt(slp=[2], icpt=[15], add=[], rem=[])
    \label{eqn:code}
\end{equation*}

The $add$ argument allows additional hits, possibly due to noise and other detector effects, to be added to the clean track, whereas the $rem$ argument removes hits from a clean track, representing the effect of detector inefficiency. The function is straightforward to implement using the equation of a straight line followed by an encoding function, as described earlier.

\subsection{Vocabulary of the new language}
From the discussions in the previous section it is clear that the new language has a very small vocabulary containing words or characters like $gen\_evt$, $slp$, $icpt$, $add$, $rem$, $($, $)$, $[$, and $]$. 

Needless to say, the vocabulary is incomplete without the numerical inputs to these arguments. Therefore, digits from 0 to 9 can be allowed as part of this vocabulary. However, from a practical point of view, only certain values for the track parameters should be allowed for the following reasons. The limited acceptance of the detectors allows only certain combinations of track parameters. For instance, a muon with a large incident angle that does not pass through all the layers may not result in a trigger.  Furthermore, due to the finite width of the strips, which dictates the spatial resolution, multiple values of slopes and intercepts may result in the same hit-ID pattern. Therefore, for this study, we have selected a set of 41,498 combinations of slope and intercepts, which ensure an optimal acceptance of the detector and a minimum of four hits in a track. 

The $add$ and $rem$ arguments are hit IDs to be added or removed, represented by integers from 0 to 383 according to the encoding format described earlier. A total of 9722 \textit{words} form the vocabulary of the language. 

In the following section, we discuss the training process and other details of translating an array of integers (hit-IDs) to our new language, in which the hits are represented as arguments of a custom function that, when executed, returns the hit-IDs.

\section{Language model architecture  and training}
\begin{figure}
    \centering
    \includegraphics[page=2, trim=5cm 0cm 5cm 0cm, clip, width=1\linewidth]{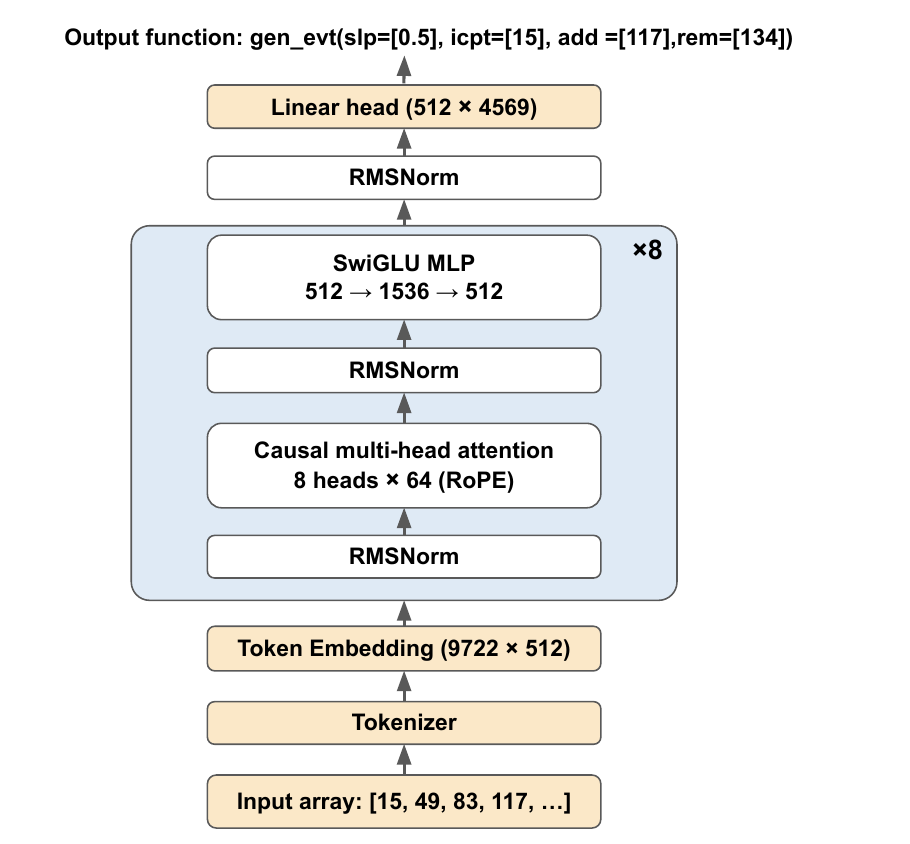}
    \caption{Architecture of our LLaMa-style small language model. An input array of integers (hit-IDs) is converted to tokens using our custom tokenizer and embedded into a 512-dimensional space. It then passes through eight LLaMA-style decoder blocks. A linear head over the 9722-token vocabulary produces the output, which is decoded into our predefined custom function.}
    \label{fig:architecture}
\end{figure}

To train this basic language model, one may choose to fine-tune open-source models. However, considering computational time available, we recognised that training a model with fewer parameters could be more expedient.
As a first step,  we designed a small decoder-only transformer model with the LLaMA architecture (LlamaForCausalLM from Hugging Face) with  8 decoder blocks, as shown in Figure \ref{fig:architecture} \cite{wolf-etal-2020-transformers}. 

\begin{figure}
    \centering
    \includegraphics[width=1\linewidth]{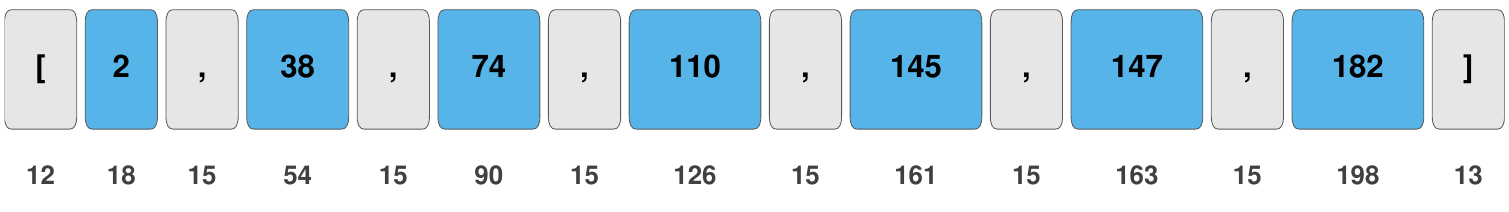}
    \caption{ Tokenisation of the input sequence via regular-expression matching with the custom vocabulary. The numbers beneath denote the token-ids.}
    \label{fig:input_tokenizer}
\end{figure}

\begin{figure}
    \centering
    \includegraphics[width=1\linewidth]{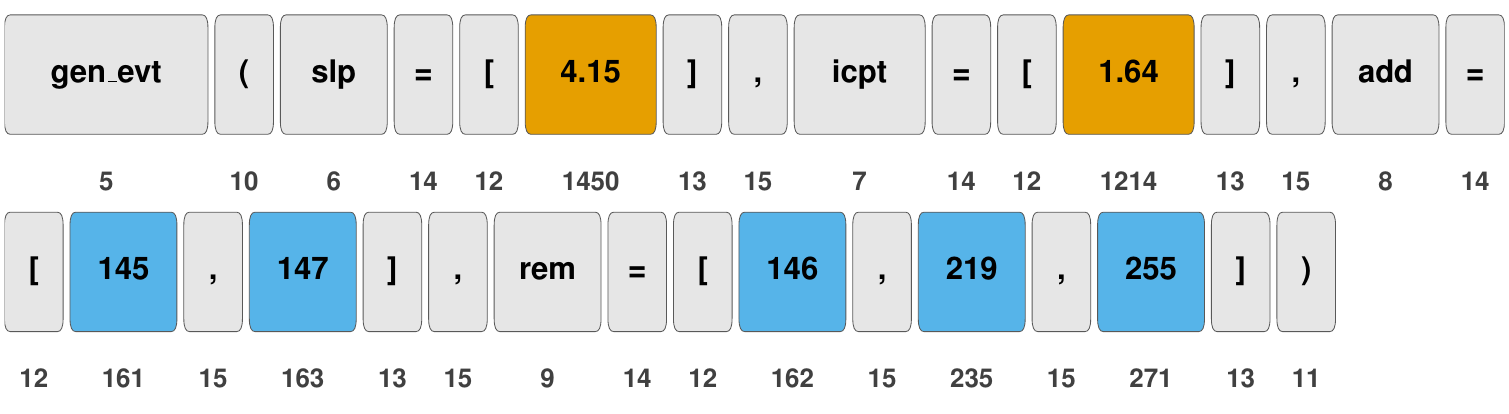}
    \caption{ Tokenisation of the output sequence in which only words that form the custom function like $gen\_evt$, $slp$ etc., are allowed in addition to the selected slope and intercept values. The arguments of $add$  and $rem$ are the usual hit-IDs.}
    \label{fig:output_tokenizer}
\end{figure}

A custom tokenizer was designed for this purpose instead of using existing general-purpose sub-word tokenizers, such as Byte-Pair encoding. 
The input sequence was tokenised using a regular expression that scans the array and extracts the hit-IDs and other syntax symbols, as shown in Figure \ref{fig:input_tokenizer}. Similarly, the output was tokenised such that only words that form the custom function, as defined in the earlier section, are allowed, in addition to the selected slope and intercept values, as shown in Figure \ref{fig:output_tokenizer}.

A custom tokenizer leads to shorter and deterministic sequences, preventing out-of-training tokens in the final output. For instance, if the tokenizer splits the numbers into individual digits, incorrect and impossible hit-IDs or slopes and intercepts may be generated, as reported in previous studies \cite{levy2025language}. Moreover, a smaller vocabulary  reduces the model size and training cost. A custom vocabulary, by allowing only certain track parameters, also serves as a medium for expressing physics or detector constraints. 

\section{Datasets}
\subsection{Synthetic datasets}
For training, testing, and validation, synthetic datasets were created by generating hits with the custom function. Hit-ID arrays, which are used as input, are produced using the 41,498 allowed track parameter combinations. In addition, noise hits are added and hits from the clean track also removed from some of the tracks. The generated events were systematically categorised to ensure comprehensive coverage of all patterns, with a distribution ratio of 1:1:1:8 among the following categories:

\begin{itemize}
    \item \textbf{Class A:} Clean
    \item \textbf{Class B:} Clean plus add only
    \item \textbf{Class C:} Clean plus rem only
    \item \textbf{Class D:} Clean plus add and rem 
\end{itemize}

The ratio exhibits a bias towards Class D events, which are more frequently represented in real cosmic events. A total of 185,000 events were produced, covering all allowed combinations of track parameters, with all four event classes represented while ensuring that the sequences of hit-IDs remain distinct across the entire dataset.

For Class B events, up to 4 extra hits were allowed, while Class D events permitted 5. In the case of Class C events, the limit for hits to be removed was established at 7, and for Class D events, it was 8. Moreover, it was confirmed that hits introduced through $add$ were not simultaneously included in the $rem$ argument.

Since generating synthetic datasets was not computationally expensive, a new set of events was created for each training epoch, using a new seed for noise and missing hits. Approximately 2,000 events were used for validation, while approximately 20,000 events were used for testing.

\subsection{Cosmic muon datasets}
Although synthetic datasets may demonstrate satisfactory model performance, the true quality of the model can only be accurately assessed using real data. Therefore, it was decided to evaluate the performance of the model using cosmic muon datasets collected from the ICAL prototype. For this test, a dataset with approximately 700,000 events was used and the performance was studied across various models and methods discussed in the next sections. Since the real datasets contain both $x$ and $y$ projections, it was possible to extract physics quantities from the model from which the performance could be understood. 
\section{Training, loss function and metrics}
The small language model was trained on a NVIDIA-H200 NVL GPU with many initial trial and error attempts regarding the choice of the architecture, the metrics and the loss functions. 
Two simple metrics were devised to assess the quality of the prediction from the model. 
The similarity index, S, denotes the fraction of hits in the input that are common to the output generated by executing the predicted function. A similarity score of 1 indicates that all the hits in the input are also present in the output.
It is also possible that the output contains more hits than those present in the input. The dissimilarity metric D quantifies this as the fraction of hits in the output that not in the input.
The metrics are thus defined as: 

\begin{equation}
  S \;=\; \frac{ H_{\mathrm{in}} \cap H_{\mathrm{gen}} }{ H_{\mathrm{in}} },
  \qquad
  D \;=\; \frac{H_{\mathrm{gen}} \setminus H_{\mathrm{in}} }{ H_{\mathrm{gen}} },
  \label{eq:sd}
\end{equation}

where, $H_{in}$ is the input sequence and $H_{gen}$ is the generated output sequence.
For an exact match between the input and output sequences, the similarity score is 1, while the dissimilarity score is 0.

One should note that the two metrics S and D alone do not represent the quality of prediction. The prediction of the correct track parameters is also an important factor.
In the first trials with a simple cross-entropy loss function, an  interesting trend was observed. Despite the inaccurate prediction of track parameters, the similarity scores remained notably high. The model accomplished this by pushing all inputs within the $add$ argument with random slope and intercept values. Based on these observations, it was necessary to modify the loss function to assign greater weight to the track parameters compared to other syntax tokens and additional arguments, while simultaneously ensuring that the length of the prediction remains optimal. As an example, in Figure \ref{fig:output_tokenizer}, the total number tokens is 32 out of which the track parameters are only 2. An incorrect  set of track parameters leads to a completely wrong prediction but an incorrect separator or syntax, though still incorrect, may be quickly learnt. Therefore, a plain unweighted loss will be dominated by trivial syntactical errors. By weighing the track parameters more than the frequently repeated symbols, the model can be forced to improve the prediction of the track parameters. The length of the prediction is also controlled by weighing the end-of-sequence token appropriately.

\section{Results}
\begin{figure*}
    \centering
    \includegraphics[width=1\linewidth]{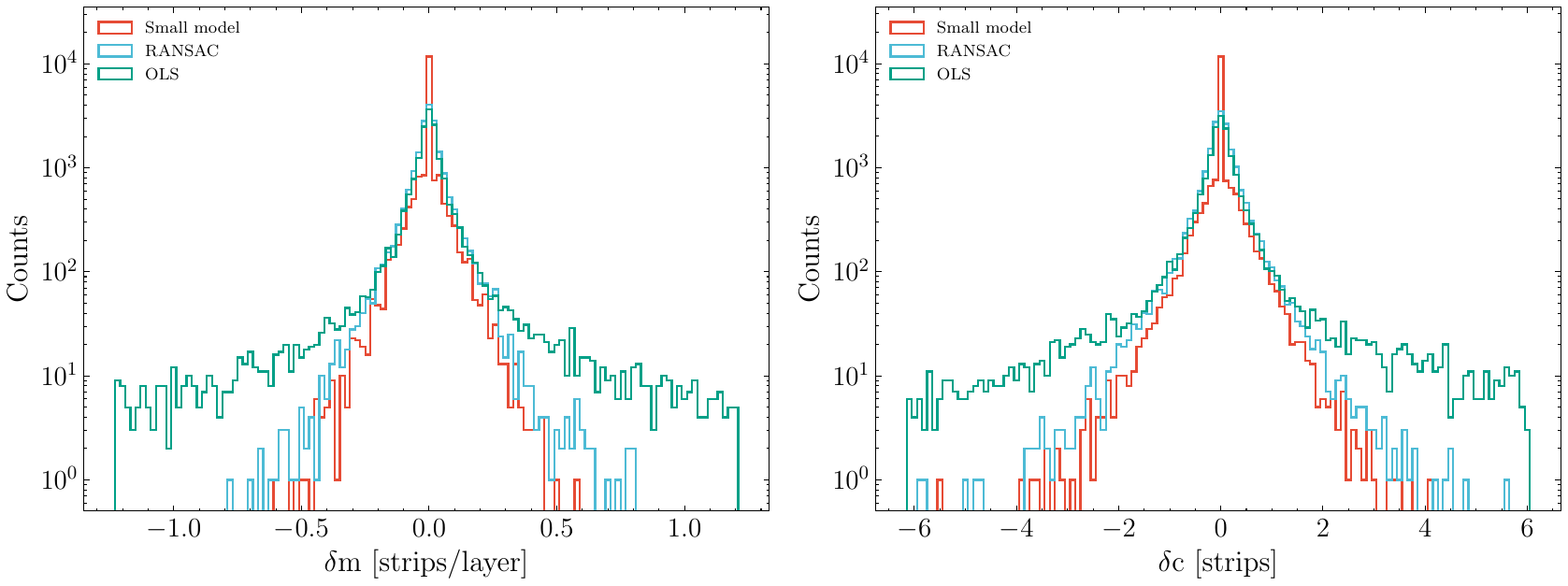}
    \caption{The residual distributions of the slope and intercept obtained from the small language model and a linear fit, clearly showing a far better performance of the model compared to the ordinary least squares (OLS) fit. For the model, only events that satisfied the cut of S>0.9 and D<0.1 (\~94\%)  are considered in this plot.
    }
    \label{fig:test_distributions}
\end{figure*}

The model was trained with an initial 500-step linear warm-up of the learning rate, followed by a cosine schedule. The training lasted for approximately 1700 epochs, taking about 16 hours.
At the end of the training, the average similarity and dissimilarity on the test dataset were 0.98 and 0.01, respectively, indicating a high quality of training. The results, summarised in Table \ref{tab:simliarity},  also indicate that while the clean tracks are almost predicted correctly, the additional noise hits impact the performance of the model as seen in the values for the perfect-fraction case.
The plots of the residual distributions of the slope $m$ and intercept $c$ are shown in Figure \ref{fig:test_distributions}. The plots clearly show the superior performance of the small model and a summary of the sample mean and standard deviations from these distributions is given in Table \ref{tab:gaussian_fit}. To minimise the influence of outliers on these estimates, a similarity cut of S>0.9 and D<1 was applied leading to a selection of 94\% of the total events. The standard deviations obtained from the model predictions are roughly a factor 10 better than the linear fit values.
Theoretical estimates for the uncertainties of the track parameters have been previously calculated and given as:
\begin{align}
    \sigma^2_c &= \frac{2N_l-1)}{6N_l(N_l+1)}&
    \sigma^2_m &= \frac{1}{N_l(N_l-1)(N_l+1)}
\end{align}

where $N_l$ is the number of layers in the detector \cite{samuel_angular_2017}. The theoretical values of the uncertainties estimated for 12 layers are also presented in Table \ref{tab:gaussian_fit} which shows that the model performance is close to ideality, even in the presence of noise.

\begin{figure*}
    \centering
    \includegraphics[width=1\linewidth]{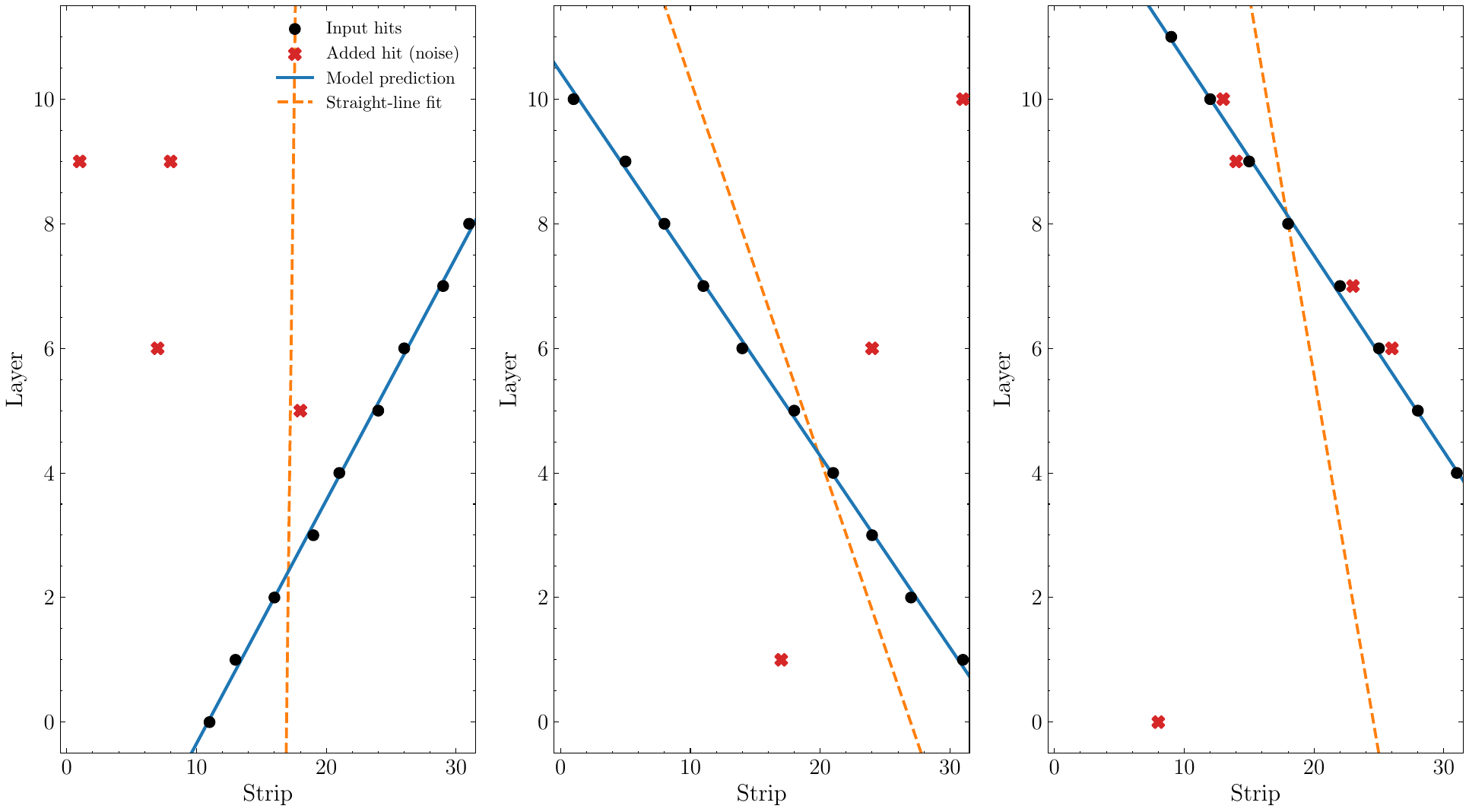}
    \caption{Sample tracks in the test dataset where the model has outperformed the conventional linear fitting which is biased by the outliers. The model is robust to noise as it is trained to jointly identify the track parameters and the noise and missing hits consistent with those parameters.}
    \label{fig:sample_events}
\end{figure*}

The performance of the model can also be understood through the sample tracks plotted in Figure \ref{fig:sample_events}, in which the model outperformed the conventional linear fit. In the presence of noise and detector inefficiencies, the linear fit, which minimises the square deviations, becomes biased, leading to larger uncertainties in the track parameters. In contrast, the model is trained to jointly identify, along the track parameters, the noise and missing hits that are consistent with the track parameters, making it more robust.   

\begin{table*}[htbp]
  \centering
  \begin{tabular}{r r r r r r}
    \toprule
    & \textbf{all} & \textbf{clean} & \textbf{add} & \textbf{rem} & \textbf{addrem} \\
    \midrule
    \textbf{mean S} & 0.98 & 0.98 & 0.97 & 0.99 & 0.99 \\
    \textbf{mean D} & 0.01 & 0.02 & 0.01 & 0.01 & 0.01 \\
    \textbf{perfect (S=1,D=0) \%} & 90.7 & 95.6 & 81.8 & 95.7 & 90.9 \\
    \textbf{exact call \%} & 53.4 & 95.1 & 59.1 & 30.2 & 18.4 \\
    \bottomrule
  \end{tabular}
  \caption{Summary of similarity scores across the four categories of events in the test dataset with 20,000 events. Exact calls are where the predicted function matches exactly with the target function.}
  \label{tab:simliarity}
\end{table*}

\newcommand{\nA}{\textcolor{red}{\textbf{212}}}
\newcommand{\nB}{\textcolor{blue}{\textbf{213}}}

\begin{table}[htbp]
\centering
\small
\begin{tabular}{p{0.46\linewidth} p{0.46\linewidth}}
\hline
\textbf{True} & \textbf{Predicted} \tabularnewline
\hline
\raggedright\ttfamily gen\_evt(slp=[\textbf{2.61}], icpt=[\textbf{4.84}], add=[144, \nB, 246, 315, 350], rem=[])
  & \raggedright\ttfamily gen\_evt(slp=[\textbf{2.63}], icpt=[\textbf{4.78}], add=[144, \nA, 246, 315, 350], rem=[]) \tabularnewline
\hline
\raggedright\ttfamily [5, 39, 74, 109, 143, 144, 178, \nA, \nB, 246, 247, 282, 315, 316, 350, 351]
  & \raggedright\ttfamily [5, 39, 74, 109, 143, 144, 178, \nA, \nB, 246, 247, 282, 315, 316, 350, 351] \tabularnewline
\hline
\end{tabular}
\caption{Demonstration of a predicted function which yields a hit-ID array similar to the true array with S=1 and D=0 but with slightly different track parameters. The true target has the hit-ID 212 as part of the clean track and the hit=ID 213 as added noise. In the predicted case, the roles are reversed.}
\label{tab:exact_match}
\end{table}
About 91\% of the predicted events in the test dataset had a exact hit-ID match, i.e, a similarity score of S=1 and D=0. 
At this juncture, it is also important to understand the reliability of the similarity scores. Specifically, for exactly matching cases, the predictions of the track parameters is expected to match the ground truth. However, it was seen that in about 40\% of such cases this was not the case. Table \ref{tab:exact_match} explains a possible cause in such predictions. For two closely matching track parameters, the noise hits can be suitably added to produce the same hit-ID sequence. The analysis of such exact-match events with wrong predictions of the track parameters showed that the  
the maximum average deviations between the clean hits of the true and predicted function is still close to 1. Even in non exact-match cases, about 6\% of the predictions matched the track parameters exactly. Consequently, these scores may serve as a metric to filter out poor predictions to some extent.

The linear fit considers all hits with equal weightage, therefore, the impact of noise is evident in the estimates of the track parameters. For a fair comparison, it is also important to consider the performance of equivalent algorithms that are not biased by outliers. The RANdom SAmple Consensus (RANSAC) algorithm is well known for its robustness against outliers \cite{fischler1981random}. The test dataset was fitted using this algorithm, and the results presented in Table \ref{tab:gaussian_fit} show that the small model was marginally better than the RANSAC analysis.

\begin{table}
    \centering
    \begin{tabular}{lllll}
    \toprule
        Type & $\mu_m $ & $\sigma_m$&$\mu_c$&$\sigma_c$\\ \toprule
         Linear fit & 0.0115 & 0.79 & -0.035& 4.87\\
         Small model& -0.0009 & 0.06    &0.005&0.40\\
         RANSAC & 0.0000 & 0.09 & 0.003 & 0.52 \\
         Theory (12 layers)& - & 0.02 &-&0.17\\
         \bottomrule
    \end{tabular}
    \caption{Summary of sample mean and standard deviations $\mu$ and $\sigma$ of the slope and intercept residual distributions, in the units of strips for the intercept and strips per layer for the slope. The theory values were calculated using an error propagation procedure, assuming 12 layers. For the small model, a similarity cut of S>0.9 and D<1 was applied after which 94\% of the events survive, using which the estimates for other models are also calculated.}
    \label{tab:gaussian_fit}
\end{table}

\subsection{Results of cosmic muon analysis}
As mentioned previously, the true test of the method lies in its performance with real data. In this study, we used the model to predict the slopes and intercepts along the $x$ and $y$ projections and then use the same to calculate the zenith angle $\theta$ distribution as given by the formula:

\begin{align}
    \theta = tan^{-1}\sqrt{m_x^2 + m_y^2}
\end{align}
The form of this distribution which depends on the physics of the process and the detector acceptance is well studied and therefore this study constitutes a true test of performance \cite{samuel_angular_2017}. 

\begin{figure}
    \centering
    \includegraphics[width=1\linewidth]{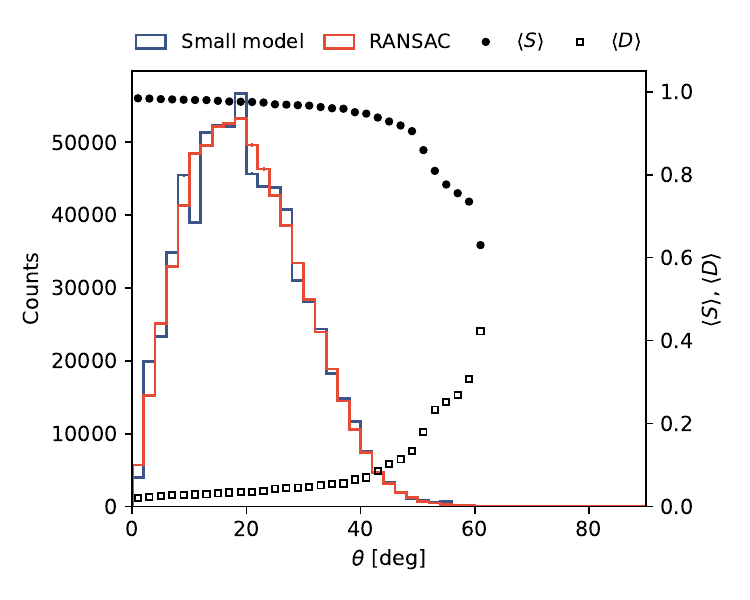}
    \caption{Comparison of zenith angle distribution using the model prediction and RANSAC along with the similarity scores from the small model.}
    \label{fig:zenith_comparison_small}
\end{figure}

The RANSAC values were used as a reference in the study, and the results are shown in Figure \ref{fig:zenith_comparison_small}. The plot shows a clear correlation between the RANSAC values and model predictions. It can also be seen in the plot that the similarity scores deteriorate at larger angles as the number of detector hits decreases, deteriorating the model ability to distinguish between noise and true hits. 
For the cosmic muon events, the mean scores were 0.97 and 0.04 for similarity and dissimilarity, respectively. The perfect match fraction was 84\% for the $x$-side and 78\% for the $y$-side, respectively, which is below the values obtained from the test set.

The study showed that the performance of the model was on par with that of the RANSAC method and that it was able to reproduce the expected zenith angle distribution from a real cosmic dataset. Therefore, while one may prefer to use the RANSAC method, the model offers few additional advantages.

One advantage is that the model achieves compression in addition to track fitting. Unlike the RANSAC method, which only provides information on the track parameters, the model returns a function that not only contains the track parameters but also deterministically reconstructs the original hits in approximately 90\% of the cases. Therefore, the generated codes are short, compressed, and complete descriptions of the original hit sequence, and their lengths correspond to the Kolmogorov complexity of the data. Therefore, the generated codes can be used as a means for efficient compression of detector data.
Although not of much use in this specific context, another advantage is the simultaneous track fitting of multi-track events. As the slope and intercept arguments are specified as arrays, multi-track events are naturally accommodated, with each pair representing a single track, as shown below:

\begin{equation*}
gen\_evt(slp=[2.1, 3.1], icpt=[15.2,17.3])
\end{equation*}
Studies on multi-muon events have been performed by many large experiments but the small size of our detector implies a negligible probability for such events \cite{acharya_multimuons_2025} \cite{samuel2024machine}. However, for the sake of completion, we present the results of extending the study to multi-track events.

\section{Studies with multi-track events}
\begin{figure*}[htpb]
    \centering
    \includegraphics[width=1\linewidth]{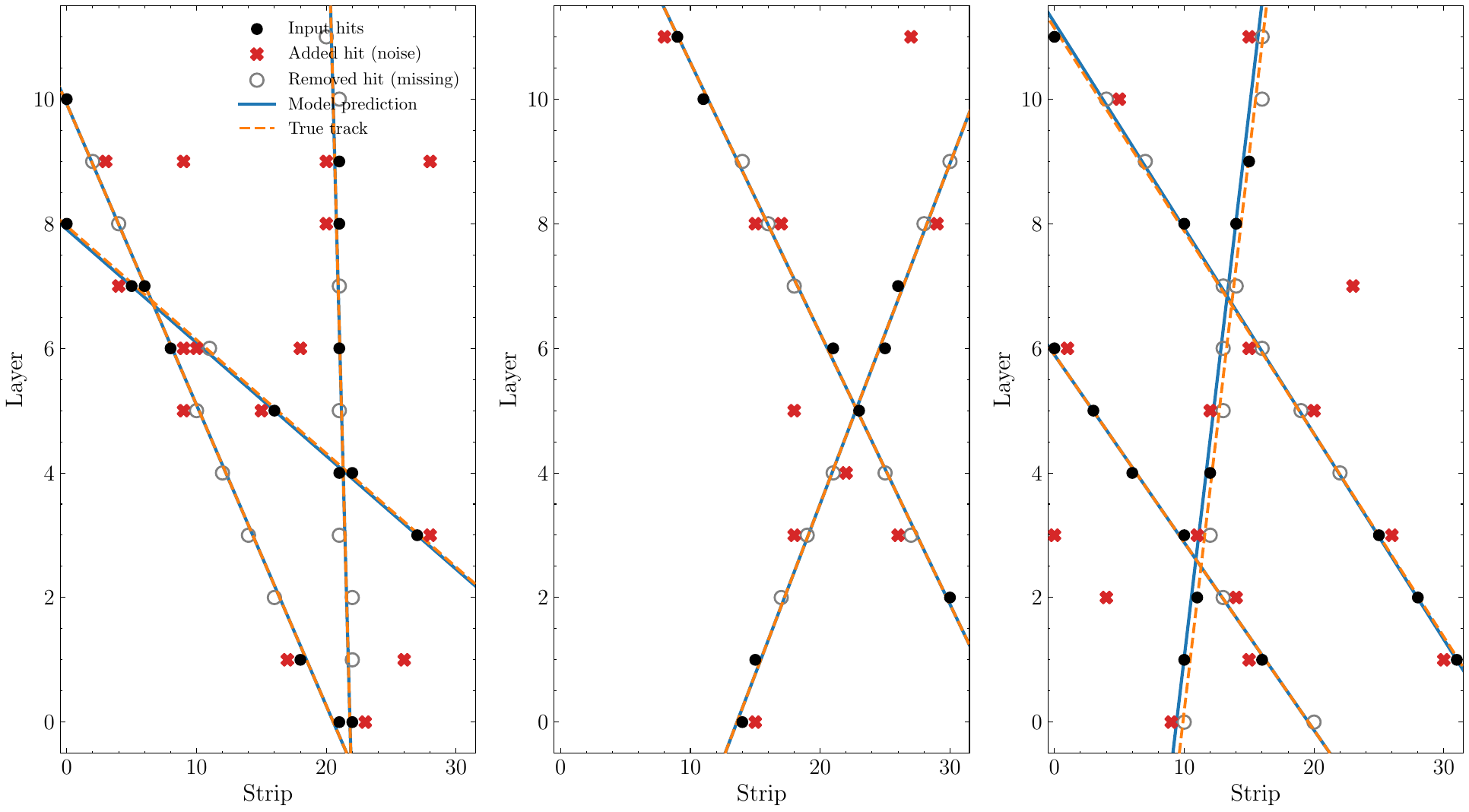}
    \caption{Few multitrack events from the test dataset correctly predicted by the large model trained on track multiplicities of 1, 2 and 3. The model also exhibits a high efficiency in track-multiplicity identification. }

    \label{fig:multi-track}
\end{figure*}

For this study, the architecture of the small model had to be enhanced to accommodate lengthier sequences. The number of layers was increased to 12, the embedding size to 1024 and the MLP to 2816 increasing the parameter size from 37.3 M to 174.2 M. Similar to the  previous model training, for each epoch approximately 120,000 events were randomly generated with track multiplicities of 1, 2 and 3 uniformly distributed. The maximum noise hits and missing hits were set to 5 and 7, proportionately increasing with track multiplicity. The training for 320 epochs lasted for about 6 h.
The results from the test dataset with 20,000 events with track multiplicities uniformly distributed across the three categories is shown in Table \ref{tab:large-extended-test}. The large model is capable of distinguishing between various multiplicities  with high efficiency with an additional improvement in the increase of perfect match fraction in the single track events to 97.1\%. However, the exact-call fraction has reduced to 36\%.
Figure \ref{fig:multi-track} shows few multi-track events correctly predicted by the model.

\begin{figure}[htbp]
    \centering
    \includegraphics[width=1\linewidth]{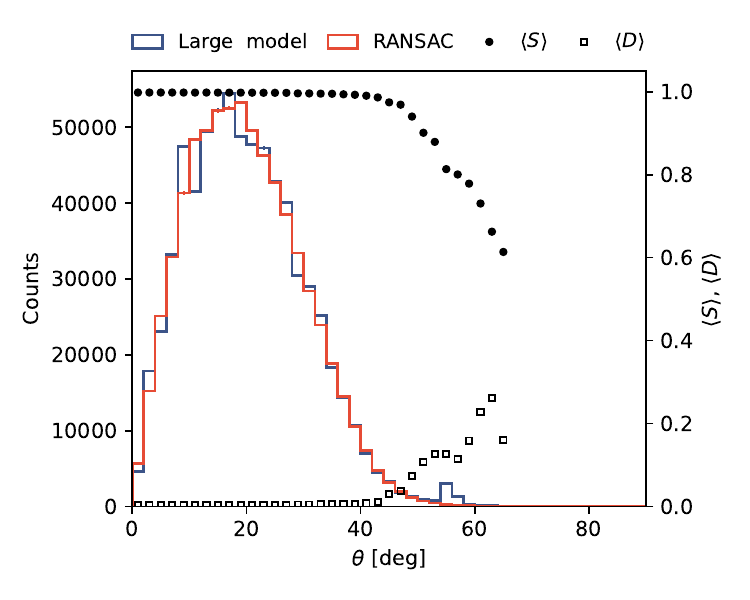}
    \caption{The zenith angle distribution as predicted by the large model. A kink close to $56^\circ$ may be attributed to steep tracks whose slopes are outside the limit in the allowed tokens. }
    \label{fig:zenith_large_model}
\end{figure}

\begin{figure*}[htpb]
    \centering
    \includegraphics[width=1\linewidth]{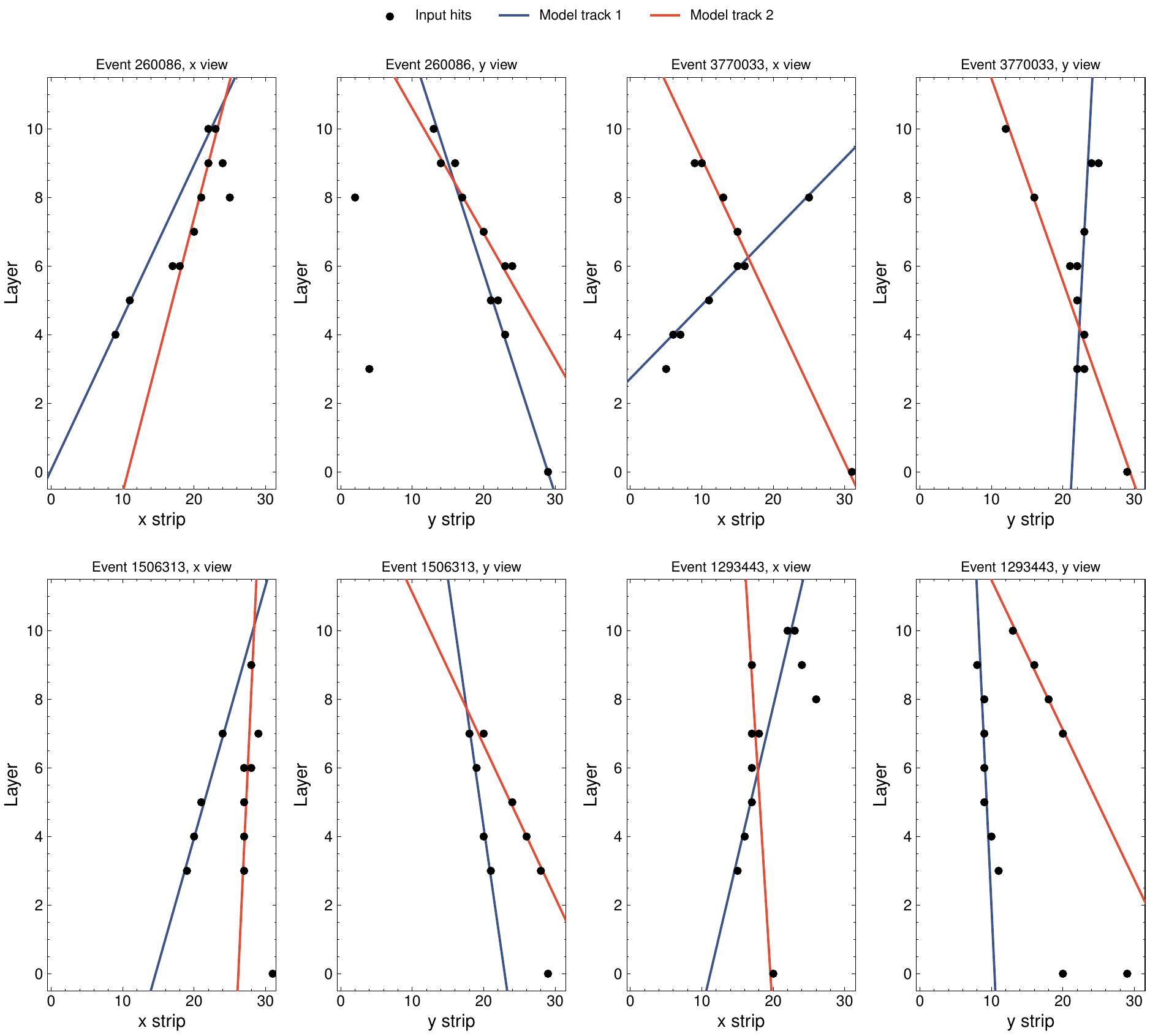}
    \caption{Few multi-track events from the cosmic muon dataset, predicted by the large model. About $5\% $ of the total events were classified with a multiplicity of 2. Shown here are the selected well separated events.}
    
    \label{fig:cosmic_multi}
\end{figure*}

The cosmic muon dataset was used to test the efficiency of the large model. The mean similarity and dissimilarity scores were 0.99 and 0.01, respectively, with a perfect fraction of 95\%. The zenith angle distribution of the large model is shown in Figure \ref{fig:zenith_large_model}. While the distribution matched well with the RANSAC values in most regions, a small kink appeared at approximately $56^\circ$ which was identified as being from steep tracks whose slope exceeds the maximum value in the allowed tokens, possibly due to noisy hits that might still lead to a trigger. In such cases, the model assigns the maximum value to the slope (8.17 strips/layer) that leads to the peak around $56^\circ$.
Approximately 95\% of the tracks were classified as single tracks, followed by 5\% of track multiplicity 2, with the rest negligibly small. However, on closer inspection, it was clear that most of the multitrack events were noisy. Some well-separated events are shown in Figure \ref{fig:cosmic_multi}.

\begin{table*}[htbp]
  \centering
  \caption{Performance of the multi-track model on the synthetic test set with 20,000 events uniformly distributed across the three multiplicities.}
  \label{tab:large-extended-test}
  \begin{tabular}{lcccc}
    \toprule
                                         & All           & 1-track       & 2-track       & 3-track       \\
    \midrule
    Perfect ($S=1$, $D=0$) (\%)          & 92.8         & 97.1         & 95.7         & 85.6         \\
    Correct track multiplicity (\%)        & 98.7         & 99.9         & 98.3         & 97.8         \\
    Mean $S$ /  $D$                  & 0.997 / 0.003 & 0.997 / 0.003 & 0.998 / 0.002 & 0.996 / 0.003 \\
    Exact call (\%)                      & 15.2         & 36.0         & 8.2          & 1.6          \\
    \bottomrule
  \end{tabular}
\end{table*}

\section{Conclusions}
In this exploratory novel work, we have presented our studies on treating track-fitting, a common algorithm in particle physics, as a language translation problem with detailed steps of the model, tokenizer and training, using a cosmic muon detector as a reference.  It is seen that the model while performing better than the ordinary least-squared fitting, yielding a 10$\times$ better accuracy, is on par with the RANSAC technique which handles outliers efficiently. With the aid of the similarity metrics, tracks with poor predictions can be identified and possibly corrected.
In terms of the tracking efficiency, at this stage, no specific benefits can be claimed of this new technique. However, unlike the usual fitting techniques which yield only the fit parameters, the model returns a executable function, which can return the original hit-sequence in 90\% of the cases. By generating a compact function whose arguments are the track parameters, the model produces a minimum-length description of the input sequence providing an upper bound to the Kolmogorov complexity. This also indicates that the output can be used for efficient compression of the detector data. 
Additionally, we have shown that the extension of this model can be used to identify multi-track events efficiently. A large model was trained with various track multiplicities which improved the fraction of perfect-match predictions to 97\%. While multi-track events may be rare for the size of the  detector considered, few multi-track events were indeed identified with the aid of the model. While track fitting in this case was straight-forward, it still remains to be seen if such models can be trained with complex detector geometries and particle trajectories, for instance, in magnetic field. Research on particle tracking in magnetic fields is currently underway, and we anticipate that it will provide a better assessment of our model's capabilities.

\section{Declarations}
\begin{itemize}
\item Funding: No funding was received for conducting this study.
\item Conflict of interest/Competing interests: The authors declares no competing interests.
\item Ethics approval and consent to participate: Not applicable
\item Consent for publication: Not applicable
\item Data availability: The data used in this work is published at \begin{itemize}
    \item \url{https://huggingface.co/datasets/deepaksamuel-cuk/trkfit-llm-datasets/}
    \item \url{https://huggingface.co/deepaksamuel-cuk/trackfit-llm-small}
    \item \url{https://huggingface.co/deepaksamuel-cuk/trackfit-llm-large}

    \end{itemize}
\item Materials availability: Not applicable
\item Code availability: The code used in this work can be found at \url{https://github.com/deepaksamuel/trackfit-llm}
\item Author contribution: D.S. conceived the study, performed part of the analysis and wrote the paper. C.E.K performed additional analysis and proof-read the paper.
\item AI/LLM use: During the preparation of this work, the author used Generative AI tools for grammar and language correction, and Claude (Anthropic) to assist in writing analysis code and generating plots/figures. After using these tools, the author reviewed, tested, and edited all outputs as needed, and take full responsibility for the content of the publication, including the validity of the analysis and results.
\end{itemize}
\bibliography{sn-bibliography}


\end{document}